\documentclass[10pt,dvips]{acta}
\usepackage{supertabular,lscape,epsfig}
\usepackage{amssymb}
\usepackage{amsmath}

\SetPages{373}{385}
\SetVol{60}{2010}
\begin{document}
\begin{Titlepage}
\Title{On the dynamical evolution of Scattered Disk Objects outside the planetary system}
\Author{R. Gabryszewski$^1$ and H. Rickman$^{1,2}$}
{$^1$Space Research Centre Polish Academy of Sciences,\\ul. Bartycka 18 A, 00-716 Warszawa, Poland\\
e-mail: r.gabryszewski@cbk.waw.pl\\
$^2$Department of Physics and Astronomy, Uppsala University,\\Box 515, Regementsv\"{a}gen 1, SE-75120 Uppsala, Sweden}

\Received{July 12th, 2010}
\end{Titlepage}

\Abstract{We report the results of dynamical simulations, covering Gyr timescales,
of fictitious Scattered Disk Objects as a follow-up to an earlier study
by Fern\'andez et al. (2004: {\it Icarus} {\bf 172}, 372). Our dynamical
model is similar in that it does not include external agents like passing
stars or the Galactic tide. Only the four giant planets are explicitly
treated as perturbers. We analyze the random-walk behavior of the
inverse semi-major axis by means of a simplified circular restricted
3-body problem as an approximate analogue. Our results concerning the
role of resonant effects and the transfer efficiency into the orbital
energy domain of the inner Oort Cloud are in broad agreement with the
earlier papers, and we confirm the important role of external objects
(with perihelia beyond Neptune's orbit) in feeding the Oort Cloud. We
estimate the efficiency of this transfer to be even somewhat higher than
previously found.}{non-resonant dynamics, Oort Cloud objects, SDO, celestial mechanics, numerical integration}

\section{Introduction}

The Scattered Disk is one of the dynamical structures identified among Trans-Neptunian 
Objects (TNOs). Its existence was confirmed by the first discoveries in the mid-nineties 
(Luu et al. 1997). The bodies of this type are generally defined by the range of perihelion 
distances and/or semi-major axes ($30 < q < 40$~AU, $a > 50$~AU). Scattered Disk Objects 
(SDOs) can be perceived as TNOs that have evolved due to perturbations at close encounters 
with Neptune (Duncan \& Levison 1997, Morbidelli \& Brown 2004). These bodies seem to be an important link between TNOs and Oort Cloud comets (Fern\'andez et al. 2004, Rickman et al. 2004).

The dynamical evolution of SDOs is also influenced by mean motion and Kozai resonances (Gomes et al. 2005, Gallardo 2006). According to these papers, objects with eccentricities $e < 0.9$ may 
experience temporary captures into mean motion resonances (MMRs). Most of these are of 
$1/N$ type with respect to Neptune, acting on Myr timescales, but occasionally Gyr 
timescales have been observed too. The Kozai resonance is often found to act on bodies in MMR 
with inclinations higher than that of Pluto. Due to oscillations in the ($e,i$) plane coupled 
to $\omega$ libration, this is a mechanism of changing the perihelion distance. Such changes 
may become permanent, if the body escapes from the resonance during a different phase of the 
cycle than where it entered.

This work presents the specific dynamics of non-resonant SDOs with large and quasi-constant 
perihelion distances outside the planetary system. Such bodies may experience large increases 
of semi-major axis and eccentricity on timescales much shorter than the age of the Solar 
System. We have analyzed the influence of orbital energy ($1/a$) perturbations by Neptune and 
other giant planets on such objects, stimulated by the work of Fern\'andez et al. (2004). 
Some remarks on the statistics of feeding the inner Oort Cloud region and the influence of 
resonances on SDO dynamics are also presented.

\section{Numerical model}

We use a sample of 100 test particles (TPs) on SDO-like orbits. The orbits are placed on a 
regular grid in ($a,q,i$) space ($i$ is the inclination), as listed in Table 1. 
The TPs cover all the grid points, and for each one we define the initial condition at the 
beginning of the integration by considering in addition a value of $\omega$ (argument of 
perihelion) that alternates between 90 and 270 degrees, and random values of $\Omega$ 
(longitude of the ascending node) and $M_o$ (mean anomaly at time zero). All the elements 
refer to the heliocentric frame. This remark is of special significance concerning the 
perihelion distances. While the pericenter distance of the barycentric orbit is relatively 
stable for high-eccentricity objects like the ones we consider, that of the heliocentric 
orbit is not. The velocity offset between the Sun and the barycenter will easily lead to 
relative changes of the perihelion distance by up to $\pm8$\%. Thus, the actual perihelion 
distances, which apply when the objects pass perihelion, may be quite different from the 
ones listed in the Table.

The initial conditions for perihelion distances, semi-major axes and inclinations were chosen 
to cover the ranges occupied by most observed SDOs, up to nearly 10~AU outside Neptune's 
orbit, and some Neptune-crossers as well. We also considered the dynamics of some particles 
with much lower starting inclinations (0 to 0.1 degrees) as a test case. The results of these 
additional integrations are not included in statistics presented in Section 3.3, but one of 
the objects forms the basis for the discussion in Sect. 3.1.

\MakeTable{c|c|c}{12.5cm}{Values of heliocentric orbital elements characterizing the initial conditions of 
test particles.}
{\hline
$a$ (AU) & $q$ (AU) & $i$ (deg.) \\
\hline
{\small 100, 200, 500, 1000}  & {\small 28, 30, 32, 34, 36} & {\small 10, 12.5, 15, 17.5, 20}\\
\hline
}

The equations of motion were integrated numerically using a hybrid method from the MERCURY 6 
package (Chambers 1999). This integrator is not fully symplectic. The MVS algorithm used in 
the symplectic part cannot handle close encounters, hence the Bulirsch-Stoer routine is used 
for integrating those approaches. The changeover distance was set to 3 Hill radii of the 
planet. 

The equations of motion were those of the $N$-body problem with the Sun, four giant planets 
and test particles. Galactic tides were not included into our Solar System model. The 
integration time step used by symplectic routines was set to $\sim$80 days. The masses of 
terrestrial planets were added to the mass of the Sun, and the TPs were assumed to be 
massless. The objects were integrated over a maximum time of 2~Gyrs, but the integrations 
were finished earlier, when one of the following conditions occurred: (a) collision with a 
planet, (b) eccentricity greater than 0.9995, (c) semi-major axis greater than $25\,000$~AU.


Various numerical tests were made before and during the integrations. 
Mainly, we inspected the outputs, comparing results from different methods of 
integration. But verification was made at nearly every step of the calculations.

\begin{figure}[htb]
\includegraphics{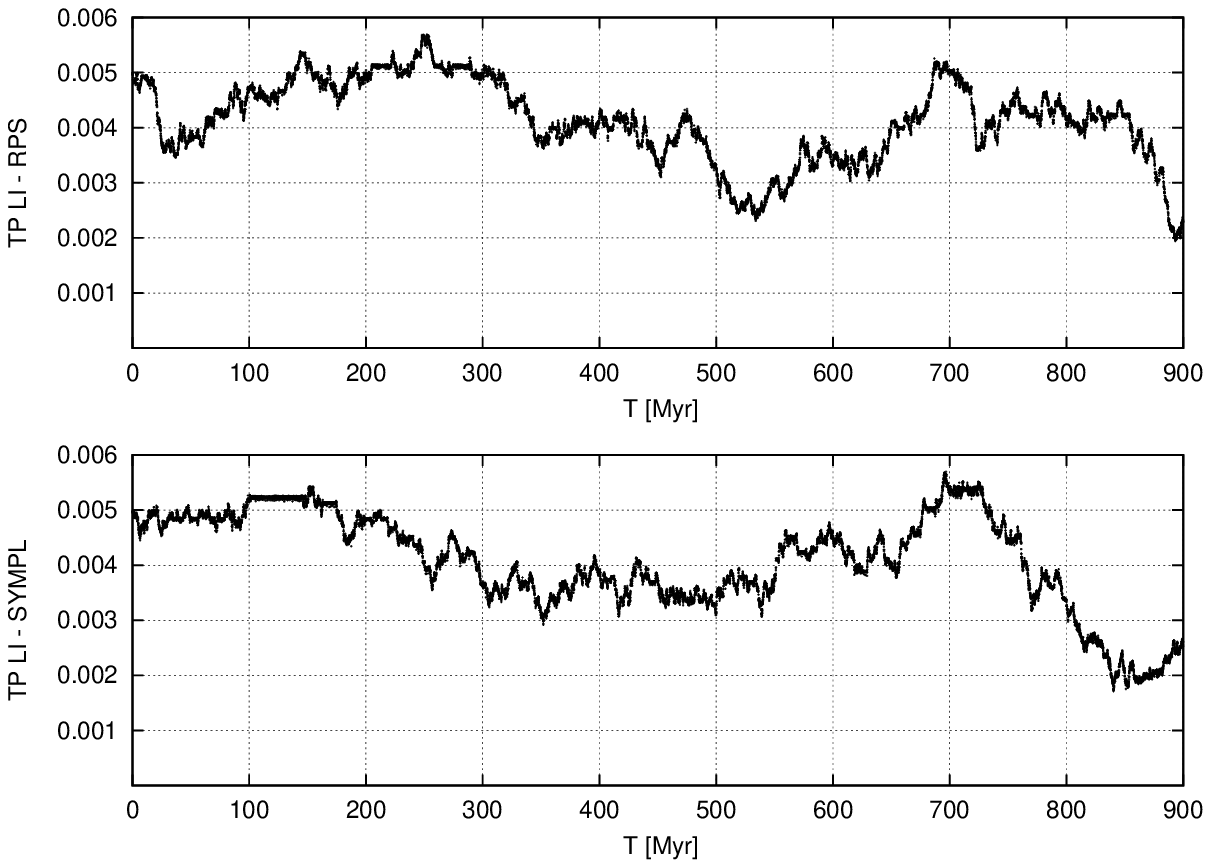} 
\FigCap{Evolutions of orbital energy $1/a$ (the inverse semi-major axis) as a 
function of time for the test particle TP~LI. The lower graph shows the 
evolution obtained by the symplectic integrator used in our calculations. The 
upper graph presents the same evolution obtained by the RPS integrator. The 
changes of orbital energies are seen to follow similar patterns in both cases. 
The unit for $1/a$ is AU$^{-1}$.}
\label{fig-1overaLI50}
\end{figure}

Besides MERCURY~6 we used the RMVS3 symplectic integrator from the SWIFT 
package (Duncan \& Levison 1997) and the recurrent power series (RPS) method 
introduced by Sitarski (2002). We examined the agreement between slowly 
varying orbital elements of test particles using all these methods until the 
first close planetary encounter. Thus, for objects without close approaches to 
the planets, we were able to verify that similar evolutions result on a time 
scale of 100~Myr. Moreover, in one case where the perihelion of the test 
particle remained more than 7.5 AU beyond Neptune's orbit (to be referred to 
below as TP~LI), we got very similar evolutions using our standard method and 
the RPS method of integration during close to 1~Gyr. This was part of a check, 
where we compared the variations of $1/a$ of randomly chosen test particles as 
calculated by the RPS method with the results from the MERCURY~6 package. The 
observed general dependencies were indeed of the same type (see Fig. ~\ref{fig-1overaLI50}).

Analysis of $1/a$ perturbations was performed using the circular restricted 3-body problem 
(CR3BP) as a simple model. A dedicated program written by Hans Rickman was used to 
calculate the perturbations $\Delta(1/a)$ experienced by an object on an unperturbed, 
parabolic orbit during one perihelion passage. This method allows to estimate the orbital 
energy change in the absence of close encounters for high eccentric SDOs evolving outside the 
planetary system with reasonable accuracy using very little computer time.

\section{Results}

\subsection{Non-resonant dynamics outside the planetary system}

Planetary perturbations acting on SDOs usually cause random variations of non-angular orbital 
parameters. This pattern can change, if the geometry of encounters is repeating on a longer 
time scale, but in the absence of such resonant behavior, the variations of eccentricity and 
semi-major axis are generally accidental, and the long-term evolution can be described as a 
random walk.

\begin{figure}[htb]
\includegraphics{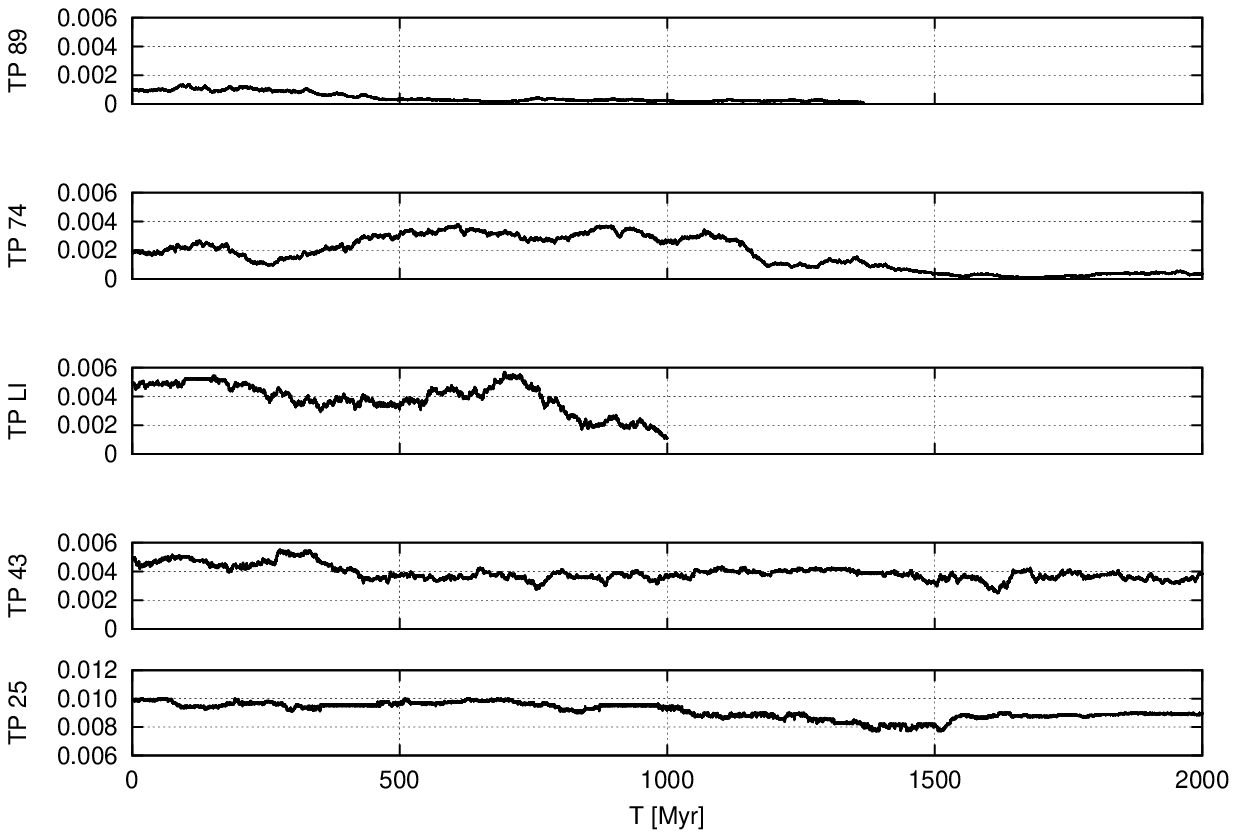} 
\FigCap{Evolutions of orbital energies $1/a$ (the inverse semi-major axis) as a function of time for five different test particles: TP~25, TP~43, TP~LI, TP~74 and TP~89. TP~LI is the particle from the test case set of initial conditions with low inclinations. It has the largest variations of $1/a$ parameter of all the bodies with quasi-constant perihelion distance. The units for $1/a$ is AU$^{-1}$.}
\label{fig-1overa}
\end{figure}

In Fig. ~\ref{fig-1overa} we show a few typical examples of the evolution of orbital energy as measured by 
$1/a$ (the inverse semi-major axis), illustrating what was said above. We see rather 
different behaviors in spite of the common random-walk picture, for various reasons. The 
case of TP~89 illustrates the behavior of particles moving into the inner core of the Oort 
Cloud with orbital periods hundreds of times larger than in the inner part of the scattered 
disk. Eventually, at $T\simeq1400$~Myr, this one is ejected from the Solar System. Other 
evolutions in the lower plots would look as flat, if their time axes were expanded by nearly 
a factor 100.

\begin{figure}[htb]
\includegraphics{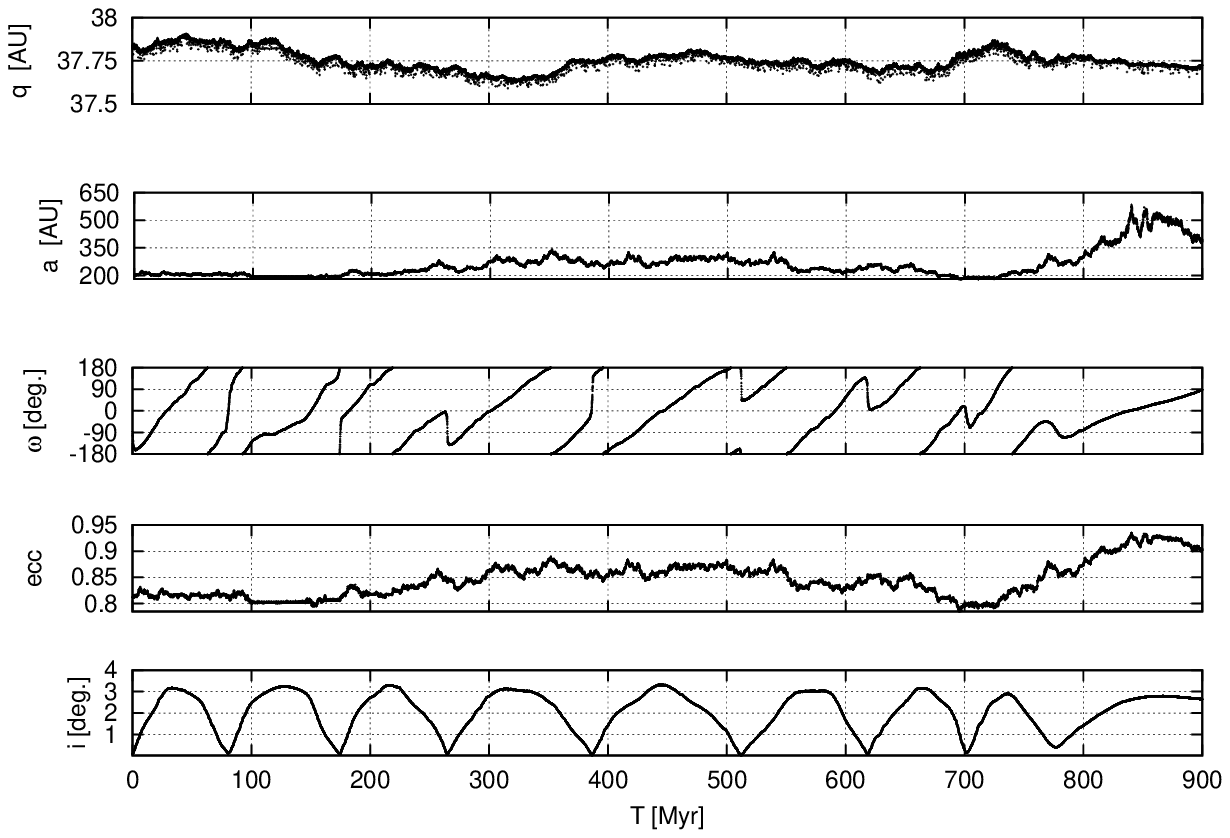} 
\FigCap{Evolutions of five barycentric elements as functions of time. Particle TP~LI was 
started with an inclination of 0.1 degree, a semi-major axis of 200~AU, and a perihelion 
distance close to 38~AU. Note that the quantity plotted here is the peribaryon distance, i.e., the 
pericenter distance of the osculating barycentric orbit. A major increase of the semi-major 
axis starts at $T\sim740$~Myr and continues for $\sim100$~Myr.}
\label{fig-1}
\end{figure}

The case of TP~25 is interesting too, and now the relative flatness of the curve is explained 
by quasi-resonant dynamics. There are many MMRs, as is typical for such small semi-major 
axes, and a few are quite long-lasting ($\sim100$~Myr). We have also included one of the 
low-inclination cases, for which the integration was limited to 1~Gyr. Particle TP~LI exhibits the 
typical random-walk behavior at first glance, but near the end there is an extended period, 
during which there is an almost steady decrease of $1/a$ ($T\simeq740-840$~Myr). Figure 
\ref{fig-1} shows some details of the dynamics of this particle.

The object is well separated from the planetary system. At perihelion it is close to 8~AU 
from Neptune's orbit, and this situation persists for the whole length of the integration. 
The change in semi-major axis between 740 and 840~Myr is seen to extend from 200 to 600 AU. 
There is no corresponding change in the behavior of the peribaryon distance, which remains 
almost constant. Instead we see a correlated, increasing trend in the eccentricity.

One interesting feature is that the variation of the argument of periapsis is distinctly 
slower compared to other time intervals -- the value remains between $-90$ and $+90$ 
degrees. Such a slow $\omega$ variation (if related to the trend in semi-major axis) may 
suggest that the limited range of $\omega$ leads to a trend in the geometry of Neptune 
encounters close to the perihelion passages. We will discuss a possible effect of this kind 
below, but we note that any $\omega$ effect must be severely limited in the present case by 
the smallness of $i$. We also investigated whether there is any repeatability of the angular 
distance between Sun-object and Sun-planet directions when the object passes perihelion, 
even though the rapidly varying semi-major axis effectively prevents any long-lasting effect 
of this kind. Indeed, no geometric repeatability was observed. 

Some insight into this kind of dynamics can be obtained using a CR3BP model in which the 
small body moves on an unperturbed parabolic orbit. A procedure for computing such `Keplerian 
estimates' of the energy perturbations $\Delta(1/a)$ has been described by Rickman (2010), 
and we follow it here, using Neptune as the perturbing planet placed at $30.1$~AU from the 
Sun. The test object has a perihelion distance of 38~AU, thus excluding the risk of close 
encounters and lending support to the approximation of an unperturbed orbit.

The integrations were done in the heliocentric frame, but we considered only the direct part 
of the perturbing function, so we actually compute the barycentric energy changes. Given the 
values of perihelion distance, inclination and argument of perihelion, only one more orbital 
parameter is needed, and we take this to be the angular distance between the projected 
Sun-object and Sun-planet lines at the time of the object's perihelion passage. Let us call 
this angle $\phi$. The closest encounters will occur near $\phi=0$. In this case the 
conjunction occurs near perihelion. Before conjunction the planet is ahead of the object, 
thus accelerating it, and after conjunction the reverse happens. Generally, for conjunction 
before perihelion, there is hence a net deceleration, and for conjunction after perihelion 
there is a net acceleration. This will be seen as maximum positive and negative values of 
$\Delta(1/a)$ occurring at either side of $\phi=0$. In case $\omega=0$, the object's 
perihelion point is practically in the planet's orbital plane, and the situation is symmetric 
so that the maximum positive and negative $\Delta(1/a)$ have the same absolute value. But if 
$\omega\neq0$, the symmetry is destroyed, and we get a predominance of either positive or 
negative $\Delta(1/a)$, depending on the sign of $\omega$. This effect will vanish for $i=0$ 
and increase in amplitude, when $i$ takes larger values.

\begin{figure}[htb]
\includegraphics{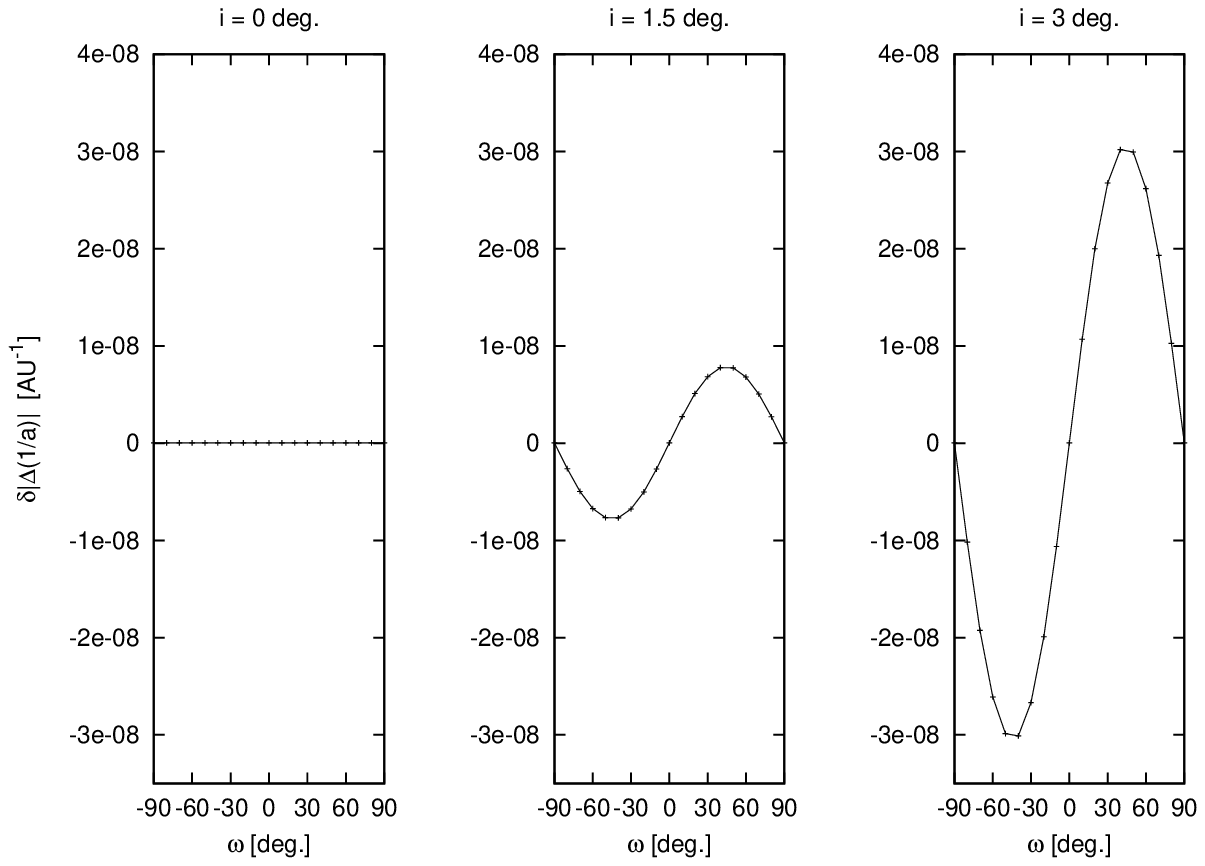} 
\FigCap{Variations of differential $1/a$ perturbations as a function of the argument of 
periapsis. The $\delta|\Delta(1/a)|$ quantity is the difference between the absolute values of the largest 
positive and negative perturbations, considering all values of the planetary longitude at 
the time pf perihelion passage of the object. The three panels give results for inclinations 
of 0, 1.5 and 3 degrees. See the text for details.}
\label{fig-4}
\end{figure}

\begin{figure}[htb]
\includegraphics{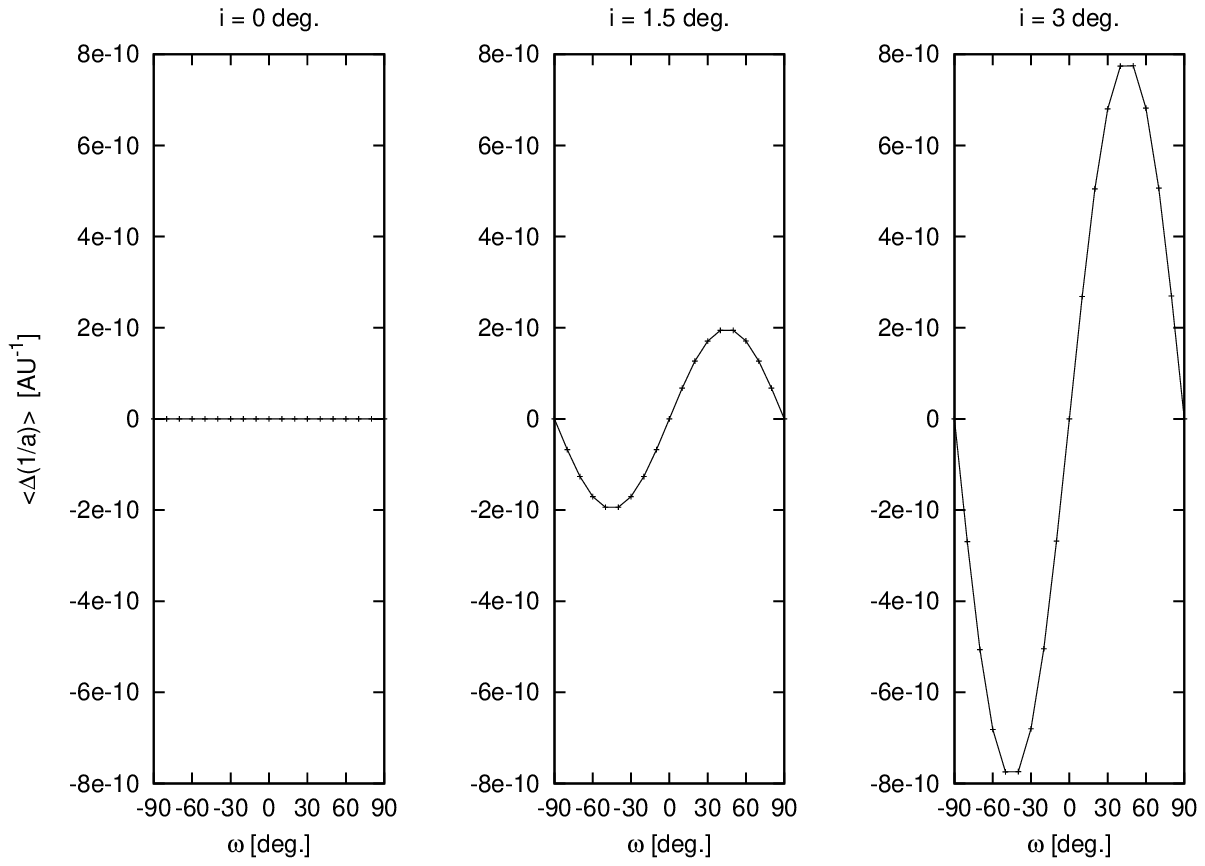}
\FigCap{Variations of mean value of $\Delta(1/a)$ perturbations taken over all values of the $\phi$ angle, as a function of argument of periapsis. The three panels show results for inclinations of 0, 1.5 and 3 degrees. Note the different scale for this quantity compared to Fig ~\ref{fig-4}.}
\label{fig-avg}
\end{figure}

Figure \ref{fig-4} shows graphs of the difference between the absolute values of maximum 
positive and negative $1/a$ perturbations vs $\omega$ for inclinations similar to the 
particle under consideration. In Fig. \ref{fig-avg} we show plots of the corresponding mean 
values of $\Delta(1/a)$, taken over all values of $\phi$. When the inclination is zero, as 
expected, the net differential or average effect is zero for all values of omega. But when 
even a small inclination is set, the differential or average $1/a$ perturbation starts to 
change in dependence of $\omega$. The largest values occur for inclinations close to 40 degrees and 
arguments of perihelion in ranges between 30 and 40 degrees in absolute value. However, they 
are far too small to have any effect in the long run, and even if they had, we would see 
opposite trends in $1/a$ depending on the sign of $\omega$, in conflict with what we observe 
in Fig. \ref{fig-1}.

In quantitative terms, we observe in Fig. \ref{fig-avg} that the average perturbation does not 
exceed $10^{-9}$~AU$^{-1}$ in absolute value for any $\omega$. A time scale of $100$~Myr 
corresponds to $\sim50\,000$ orbital revolutions, and thus, even if $\omega$ would be locked 
at the optimum value for this period of time, the cumulative perturbation would be less than 
$5\cdot10^{-5}$~AU$^{-1}$, while the change observed in Fig. \ref{fig-1} is about 0.003 in 
the same units. On the other hand, the general level of the $1/a$ perturbation at an 
arbitrary value of $\phi$ is $\sim1\cdot10^{-5}$~AU$^{-1}$, so a random walk with $50\,000$ 
steps typically covers a range of 0.002 in rough agreement with what we see in the Figure.

Note, however, that high-inclination objects stand a better chance of being influenced by the considered differential effect. Close to $i = 40^\circ$ we find average perturbations reaching close to $10^{-7}$ AU$^{-1}$ in some ranges of 
$\omega$. Thus, if $\omega$ circulates very slowly, significant evolution of $1/a$ may in fact occur at such high inclinations.

\subsection{Resonances and quasi-resonant states}

During their evolution SDOs can experience temporary captures into two types of resonance: mean motion (MMR) and Kozai type. Due to the weak planetary perturbations outside Neptune's orbit, high-order MMRs may have a significant influence of SDO dynamics. This observation is in agreement with the conclusions of Gomes and Gallardo, but there are also some differences. 

In our results the maximum time intervals of single resonant states are up to $\sim100$~Myr and most often they are close to $20 - 30$ Myr. This means the timescales are generally shorter than Gomes' results, but this difference may be caused by distinctions of the considered orbital populations. MMRs of $2/N$ type were as frequently observed as $1/N$ type. MMRs of $3/N$ type were noticed only for SDOs evolving toward small semi-major axes. The time intervals of this latter resonant type were much shorter than in the other mentioned cases. All long lasting resonant states occurred with Neptune, and no stable MMRs were observed with any other planet. 

\begin{figure}[htb]
\includegraphics{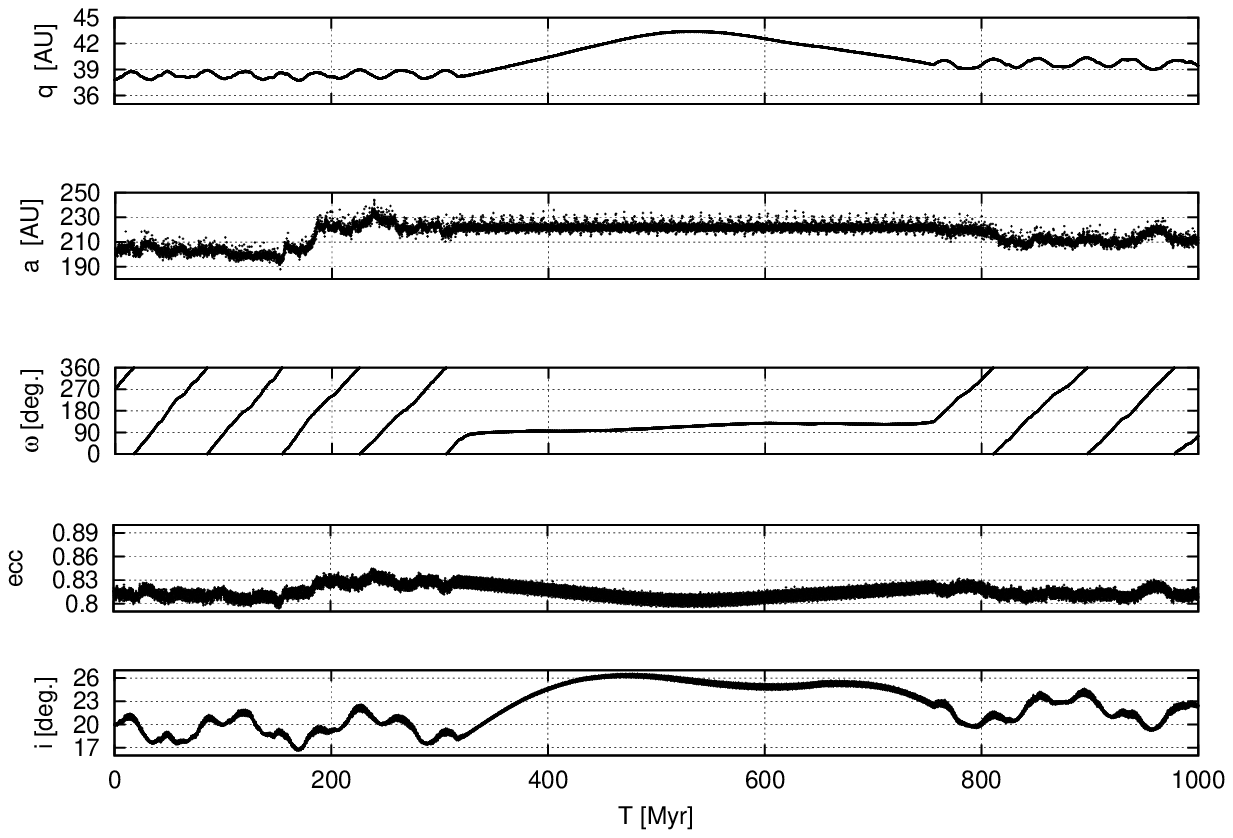}
\FigCap{TP 50: Dynamics of a body under the influence of the Kozai resonance. Its action 
changes the perihelion distance by only about 2~AU.}
\label{fig-8}
\end{figure}

Kozai cycles were observed in the simulations far less frequently than MMRs. We noticed only 
three of these. The longest capture lasted 400~Myrs (see Fig. \ref{fig-8}), and none of them were coupled with any 
MMR. In all cases the change of perihelion distance was in the range of 2~AU or smaller, 
which is to be expected since we do not consider any very high inclinations. We did not 
observe the interruption of any Kozai cycle in a different phase than the one where it was 
entered. It seems that when the Kozai resonance is not associated to a MMR, there is no 
mechanism able to interrupt it in such a way as to trigger a large change in 
perihelion distance. This suggests that the Kozai resonance as a mechanism of perihelion 
distance increase need not be very efficient.

\subsection{Oort Cloud transfers}

\begin{figure}[htb]
\includegraphics{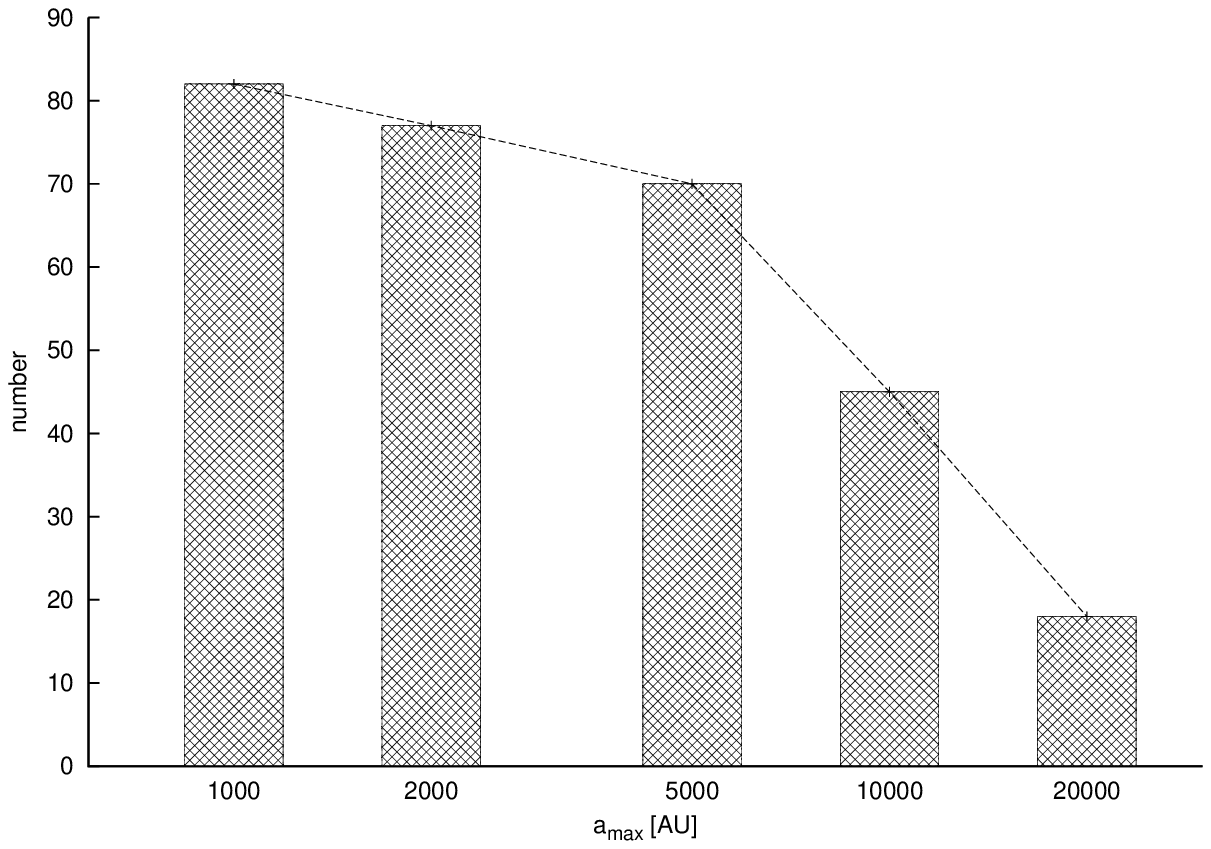}
\FigCap{Histogram of semi-major axis maximum values (plotted on a log 
scale) for our sample of SDOs.}
\label{fig-dist}
\end{figure}

The simulations allowed us to estimate how many SDOs may reach the 
inner Oort Cloud within the given time. Figure~\ref{fig-dist} shows the 
cumulative distribution of maximum semi-major axes for our 100 objects 
sample. Most SDOs evolve into the range of $1000 < a < 10\,000$~AU, 
but the number of objects breaking the $10\,000$~AU limit is much lower. 
If we assume an inner Oort Cloud limit at 5000~AU, the fraction of objects 
entering this region is 70\%. If the limit is set to $10\,000$~AU, the 
fraction of evolving objects decreases to 45\%.

These results were obtained by counting the number of objects with 
semi-major axes taking values larger than the limit. The temporary 
acceleration and deceleration of an object encountering Neptune to 
within three Hill radii may cause rapid changes of $1/a$ that are very 
important for highly eccentric SDOs like those we are investigating.
Thus we might register very large values of the semi-major axis of a 
SDO close to the moment of its perihelion passage, when a close approach 
might occur. To avoid such errors, the condition if the semi-major axis 
surpasses the given limit was verified at true anomalies of more than 
$1^\circ$ after perihelion when changes of the semi-major axis in subsequent 
data dumps were always $<5$~AU.

Our obtained numbers, quoted above, are in a good agreement with the 
outcomes of Fern\'andez et al. (2004), who used a different condition, 
namely, the penetration of the object beyond $r=20\,000$~AU in heliocentric 
distance. The difference in fractions of objects reaching an inner Oort 
Cloud starting at $a=10\,000$~AU is only about 5\%. Note, however, that 
the timescales of the two investigations are different: 2~Gyr in our case, 
and 5~Gyr for Fern\'andez et al. (2004). We can therefore expect that our 
percentage would become a bit larger than that of Fern\'andez et al. (2004), if we 
had integrated over as long a timescale as they did. Nonetheless, the 
difference is remarkably small, noting the large variations of initial 
conditions. The population used by Fern\'andez et al. (2004) is based on 76 
observed bodies with inclusion of their clones, while in this work we used 
a distribution of arbitrarily chosen initial conditions. Additionally, half 
of these objects had much larger values of initial semi-major axes (500 and 
1000~AU), comparing to the population used by Fern\'andez et al. (2004).

Let us note that limit of $a = 10\,000$ AU is rather conservatively estimated. Many authors, like Duncan et al. (1987), advocate a closer limit for external agents to remove the perihelia -- closer to the above number of 5000 AU. Thus we may estimate that a large majority of our treated SDOs will actually become potential inner Oort Cloud bodies on a 2 Gyr timescale. 

\begin{figure}[htb]
\includegraphics{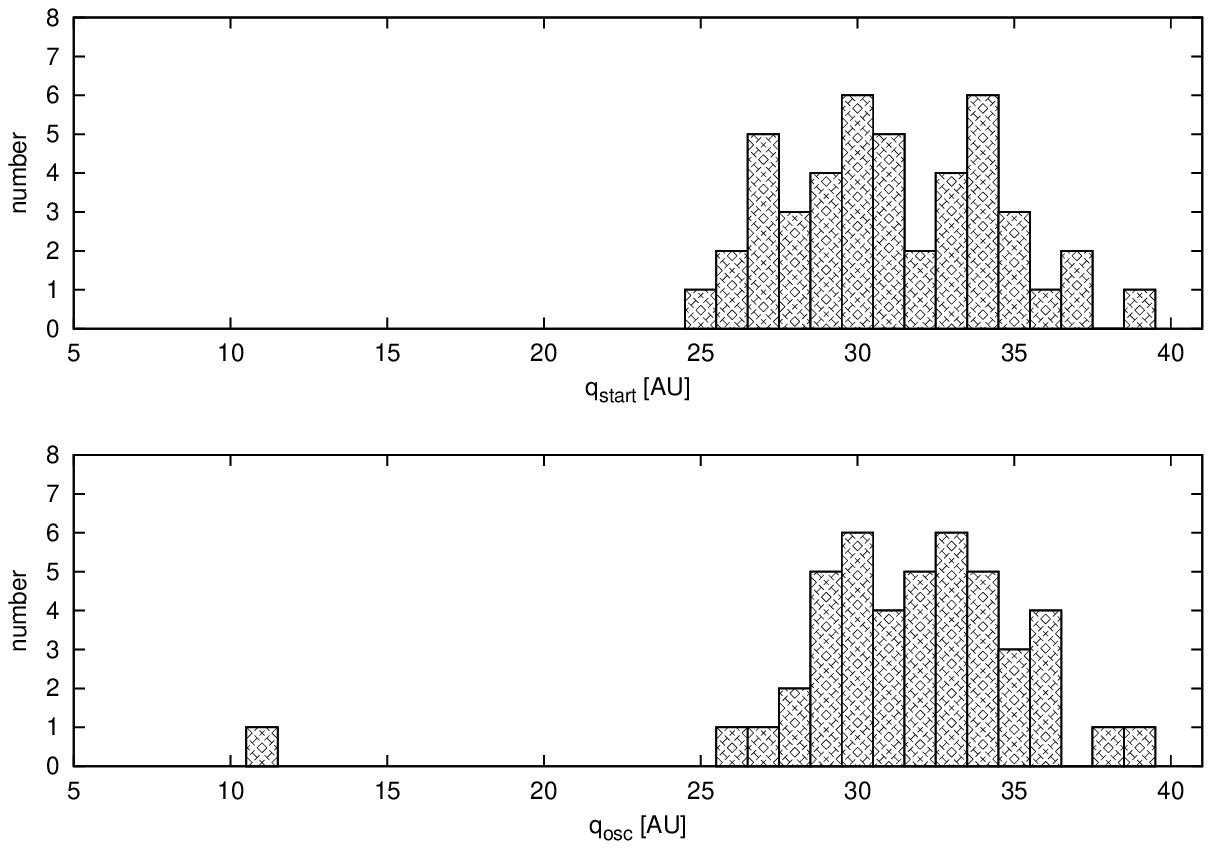}
\FigCap{Number of SDOs reaching the inner Oort Cloud (assumed to start at 
$a=10\,000$~AU) as a function of starting peribaryon distance (upper graph) 
and osculating peribaryon distance at the time of entry (lower graph).}
\label{fig-3}
\end{figure}

While the fraction of SDOs entering inner Oort Cloud is thus similar 
to the one of Fern\'andez et al. (2004), we find some differences in the 
distributions of elements upon entry. Figure~\ref{fig-3} presents the 
distributions of perihelion distances (or rather peribaryon due to the use 
of barycentric reference frame) of bodies that reached an inner Oort Cloud 
starting at $a=10\,000$~AU -- the initial distribution on the upper graph 
and the distribution at the moment of entry on the lower graph. The graphs 
are mutually similar, indicating very weak variations of the peribaryon 
distances of SDOs on their way to the inner Oort Cloud.

Planetary perturbations change the orbital energies of SDOs very 
efficiently and transfer them into different Solar System regions, but their 
peribaryon distances are much less diffused outside Neptune's orbit. This 
is characteristic of low-inclination evolution within the CR3BP. But as 
the lower panel shows, objects with perihelion/peribaryon distance over 
36~AU can enter the inner Oort Cloud due to sole planetary perturbations 
within a 2~Gyr timescale. Even if we convert the values to the heliocentric 
reference frame, we get statistically the same result. Fern\'andez et al. (2004) 
did not obtain this even after their 5~Gyr simulation. However, they 
did get more transfers caused by Jupiter, Saturn and Uranus than we did. 
We may view this as an independent confirmation of the effect they called 
Neptune's dynamical barrier, favoring outward evolution over the other 
end states.

It is hard to compare our results with the ones of Leto et al. (2009) due 
to differences of the Solar System model and the definition 
of the inner Oort Cloud region. Leto et al. started their integrations 
from quasi-circular orbits in the protoplanetary disk with the inclusion of 
perturbations due to the giant planets, the Galactic tides and stellar 
encounters. They assumed that an object reached the Oort Cloud, when 
$q > 50$~AU and $2000 < a < 25\,000$~AU. In the presence of such large 
differences, no correspondence of outcomes should be expected.

\section{Conclusions}

Conclusions from the results can be summarized as follows:

\begin{itemize}

\item Planetary perturbations are able to significantly change the orbital 
energies of SDOs, even if these are detached from the planetary system. The 
perihelion distance of the objects reaching inner Oort Cloud may remain 
quasi-constant on dynamically long timescales. The influence of very 
slow variations of the argument of periapsis via differential $1/a$ 
perturbations is far too small to explain the observed large changes in 
semi-major axis for low-inclined particles.

\item The influence of $\omega$-related differential $1/a$ perturbations 
depends on the inclinations of the orbits. The larger the inclination of 
an orbit, the larger these perturbation become, with a maximum over 
$i\simeq40^\circ$. The largest value of differential $1/a$ perturbations 
are found for $\omega\simeq 30-40^\circ$ in absolute value.

\item The Kozai resonance does not seem to be a very efficient mechanism 
of increasing perihelion distances, unless it is associated to a MMR.

\item Diffusion of perihelion distances of SDOs is low during 2~Gyr 
evolution outside Neptune's orbit. But objects with perihelion/peribaryon 
distances larger than 36~AU are able to enter the inner Oort Cloud region 
within this timescale.

\item As noted by Fern\'andez et al. (2004), SDOs can be perceived as a source of 
feeding the inner Oort Cloud population. Our results indicate that large 
fractions of SDOs with perihelia in the range out to at least $\sim35$~AU 
do enter at least temporarily into the semi-major axis range of the inner 
Oort Cloud. They may form the basis of simulation techniques, whereby we 
may efficiently investigate the outcomes of including the Galactic tides and 
stellar encounters. Thus we may in the future be able to check the efficiency
of actual feeding, as estimated by previous authors (e.g., Leto et al. 2009), 
by removal of the perihelia beyond the range where the planets act.


\end{itemize}

\Acknow{We thank the referee for helpful comments that improved the manuscript. This work was supported by the MNII grant N N203 392 734.}

\end{document}